\documentclass[runningheads]{llncs}
\usepackage[T1]{fontenc}
\usepackage[inline]{enumitem}
\usepackage{xcolor}
\usepackage{graphicx}
\usepackage{booktabs}
\usepackage{hyperref}
\usepackage{array}
\usepackage{orcidlink}

\begin{document}
\title{Data Discovery using LLMs - A Study of Data User Behaviour}
\titlerunning{Data Discovery using LLMs}
%
\author{Christin Katharina Kreutz\inst{1,2}\orcidlink{0000-0002-5075-7699} \and
Anja Perry\inst{3}\orcidlink{0000-0003-0574-9275} \and
Tanja Friedrich\inst{4}\orcidlink{0000-0003-1557-3728}}

\authorrunning{C.K. Kreutz et al.}
%
\institute{TH Mittelhessen - University of Applied Sciences, Gießen, Germany \\
\and 
Herder Institute, Marburg, Germany \and 
GESIS - Leibniz Institute for the Social Sciences, Cologne, Germany \and 
German Aerospace Center, Cologne, Germany}

\maketitle              
\begin{abstract}

Data search for scientific research is more complex than a simple web search. 
The emergence of large language models (LLMs) and their applicability for scientific tasks offers new opportunities for researchers who are looking for data, e.g., to freely express their data needs instead of fitting them into restrictions of data catalogues and portals. However, this also creates uncertainty about whether LLMs are suitable for this task.
To answer this question, we conducted a user study with 32 researchers. We qualitatively and quantitively analysed participants' information interaction behaviour while searching for data using LLMs in two data search tasks, one in which we prompted the LLM to behave as a persona.
We found that participants interact with LLMs in natural language, but LLMs remain a tool for them rather than an equal conversational partner. This changes slightly when the LLM is prompted to behave as a persona, but the prompting only affects participants' user experience when they are already experienced in LLM use.

\end{abstract}
%
%
%



%
\section{Introduction}

Finding data for scientific research continues to be a major challenge~\cite{10.1145/3529372.3530931}. To ensure that research data can be shared and found online, data producers would have to provide adequate description
~\cite{10.1007/978-3-031-16802-4_55}, ideally in a standardised form. Unfortunately, this is not the case for many research data produced today~\cite{skopal_improving_2020}. Understandably, scientists have established their own workarounds for finding data. Studies show that they learn about 
data through engaging with a topic: either from papers that cite relevant data or from exchange with other researchers, e.g., during their academic education, in their research group, or at conferences~\cite{friedrich_looking_2020,Gregory2019b}. Oftentimes, finding data resembles an iterative, collaborative process~\cite{hulsebos_it_2024} rather than a web search or a simple keyword query.

In addition, even designated systems for data retrieval (e.g.,~\cite{10.1007/978-3-030-30760-8_39,CrespoGarrido2025,dspace}) do not meet the needs of researchers 
looking for data. They usually do not support flexibility in search requests, but rather impose query restrictions on their users~\cite{https://doi.org/10.1002/pra2.457}. When given the opportunity to freely phrase requests, users of data catalogues or portals use conversational natural language~\cite{Kacprzak2018}, which raises the question whether the recent widespread availability of large language models (LLM) offers the opportunity to engage with data sources more easily. 
Conversational search via LLMs such as ChatGPT or Perplexity overshadows the use of conventional search engines in daily information seeking as well as in professional settings. Is this also an opportunity to meet the challenge of finding research data? In order to 
answer this question, we designed a study to investigate \textit{the information seeking practices of researchers looking for research data using LLMs}.

From a total of 32 participants from different disciplines, we collected survey responses as well as observational and interview data regarding their efforts to find research data using ChatGPT or Perplexity. This paper describes our \href{https://github.com/kreutzch/TPDL25_DataDiscoveryUsingLLMs/blob/main/Research%20Design.pdf}{study design}, interaction data~\cite{ZA8800} and code\footnote{\href{https://github.com/kreutzch/TPDL25_DataDiscoveryUsingLLMs}{Code for analysis of dataset and research design on GitHub}.}, and analyses of our collected data with regard to \textit{i)} the observable interaction behaviour of participants and LLMs, \textit{ii)} humanisation, and \textit{iii)} differences in user experience between user groups.

\section{Related Work}

In recent years, studies have investigated data seeking, data search, or data discovery in different contexts and from different perspectives.
Strategies employed for data search have been found to be barely distinguishable from the strategies for paper search based on used interaction paths~\cite{DBLP:conf/ercimdl/CarevicR020}. Nevertheless, searching data is assumed to be more complex than paper search~\cite{10.1145/3529372.3530931}. Simple keyword search does not cover data seekers' information needs~\cite{DBLP:journals/vldb/ChapmanSKKIKG20,Kacprzak2019} and it has been suggested that difficulties in data retrieval from information systems are linked to the query restrictions that these systems impose on users~\cite{https://doi.org/10.1002/pra2.457}. Data users are inclined to phrase their searches according to what they believe the system can process, for example, a keyword search~\cite{https://doi.org/10.1002/pra2.457}. Specifics and vague requirements (such as time spans) are lost in this process~\cite{https://doi.org/10.1002/pra2.457}. Data seekers may employ a mixture of strategies and combine multiple resources to find data, e.g., a Google search to locate repositories. However, the success of this strategy is mixed~\cite{Gregory2019b}.

Apparently, data search goes beyond simple queries and can more aptly be described as a "complex socio-technical process"~\cite{Gregory2019b} involving various aspects, such as reading literature and searching databases, but also referring to personal networks, collaborations, or visiting conferences~\cite{Gregory2019b}. It is common that data are found serendipitously~\cite{hulsebos_it_2024}, and again, through reading papers or having informal conversations with colleagues: "dataset search is not just a query, but an iterative and collaborative process involving many humans-in-the-loop"~\cite{hulsebos_it_2024}. Specifically, for frequent formal or informal exchanges regarding data search and access within a scientific community, the term "data talk" has been coined~\cite{DBLP:conf/asist/Yoon17}.
 
The use of conversational language in dataset queries has also been studied more closely. A study of crowdworkers showed that their natural language requests had a higher word count than common queries in retrieval systems, the word count being more similar to "question queries" than keyword queries~\cite{Kacprzak2018}. This finding suggests that, if given the opportunity, people looking for data could increasingly employ "longer queries, that are closer to natural language"~\cite[p.~13]{Kacprzak2019}. This finding is just one step away from using LLMs instead of search engines to find data. We already see that many researchers use LLMs as a research tool, most of them for information seeking and editing tasks~\cite{liao2024llmsresearchtoolslarge}.

\section{User Study}
Going beyond previous work, this study investigates the search for data in the timely setting of using LLMs. Specifically, we conducted a study in which we examine and compare data search behaviour both via free interaction with an LLM and via a simulated "data talk" with an LLM. We consider two scenarios: one in which study participants are free to use the LLM however they see fit ("unguided task") and one scenario, where we use the effective technique~\cite{white2023promptpatterncatalogenhance} of pre-prompting the LLM to take on a specific persona, mimicking a colleague for data talk ("guided task").








\subsection{Study Setup}

In February and March 2025, a trained instructor conducted one-on-one interviews~\cite{ZA8800} in English with each participant via a Zoom call with the help of a questionnaire in LimeSurvey~\cite{LimeSurvey}. 
Prior to conducting the study, ethics approval was received from the German Aerospace Center, and three pretests have been conducted.

\begin{table}[t]
    \caption{Task descriptions given to participants and prompt entered into LLMs.}

\tiny
    \centering
    \begin{tabular}{lp{10cm}}
        \textbf{Step II:}& Imagine you are going on a trip to Heidelberg. Use the LLM to find an interesting place to visit based on your interests. Think aloud as you explore options and explain your final choice.\\
        \textbf{Step III:}& You want to collect new data to answer your current research question. As a first step, you now look to see whether data of this type already exists and whether you can answer your research question with existing data.\\
        \textbf{Step V:}& You now have a new research idea and need to collect data for it. You first want to see whether suitable data already exists elsewhere. For this task, imagine you are not chatting with an LLM but with a colleague.

        Imagine, you start a chat via a messenger with your colleague Leslie who you know has a very good overview of various research data in your field of research. Start by explaining your data need by describing which questions or variables you are looking for.\\
        \textbf{LLM prompt:}& For the following chat, I want you to act as this character: Your name is Leslie and you are my coworker. Conduct the conversation with me from the point of view of Leslie. Do not break character and never flip roles with me. I will be asking you for help regarding the search of a scientific dataset that I want to work with\\
    \end{tabular}
    \label{tab:prompts}
\end{table}

\textbf{Step I: Rights and Introduction.}
Before a session, participants received the informed consent form\footnote{\href{https://github.com/kreutzch/TPDL25_DataDiscoveryUsingLLMs/blob/main/DataProtectionInformation.pdf}{Data protection information on GitHub}.} by email that included the general study procedure. At the beginning of the session, the participant was welcomed and again informed that the whole session would be audio and screen recorded\footnote{The recordings will be transcribed using aTrain~\cite{haberl2023atrainintroducinginterfaceaccessible} and analysed at a later date.}. After the participants verbally gave their informed consent to participate, the recording started. 

Each participant was assigned an LLMs (see \autoref{sec:llms}). The interviewer explained to the participant how to interact with the LLM through the video conference system. This ensured that participants did not need own accounts for an LLM and that the search history was accessible to the interviewer. 

Some initial questions targeted participants' primary research fields, research topics, the number of times and tasks for which LLMs have already been used for academic tasks, and if an LLM would be used for searching for research data.

\textbf{Step II: Pre-Study Training.}
To become familiar with the concept of thinking aloud and using the assigned LLM remotely, we asked study participants to perform a small practice task (see \autoref{tab:prompts}) while thinking aloud\footnote{Concurrent think-aloud provides insights into cognitive processes in the working memory without interfering with the task performed~\cite{Ericsson1980} and has been used in various user experience studies~\cite{JASPERS2004,DBLP:conf/jcdl/KreutzBSSW23,DBLP:journals/corr/abs-2403-15861}. The data will be analysed at a later point in time.}. 
The interviewer instructed the participants to speak continuously, to say everything that comes to mind, related to the task or not~\cite{Noushad2023}. In general, during think aloud phases, the interviewer remained silent and only reminded participants to think aloud in case they forgot.
%
The participants were stopped after working on the task for three minutes.

\textbf{Step III: Unguided Task.}
The first data search task (see \autoref{tab:prompts}) using the assigned LLM is performed without instructions on how to conduct the search itself, except for thinking aloud. After seven minutes, the participant's search was stopped by the investigator in case no earlier stopping decision was made.

\textbf{Step IV: Experience with Unguided Task.}
After completing the unguided task, participants filled out two user experience surveys (the User Experience Questionnaire~\cite{UEQ}\footnote{Without the \texttt{Novelty} subscale.} and the NASA Task Load Index~\cite{hart1988development}) in the questionnaire. Following that, a short verbal interview asked participants to describe the liked and disliked aspects of the unguided task, the familiarity of the data searched for, and how the task would have been solved, if they were not asked to use the assigned LLM.

\textbf{Step V: Guided Task.}
The second data search task (see \autoref{tab:prompts}) was performed with the assigned LLM that was prompted beforehand to behave as a persona mimicking a helpful colleague (see \autoref{tab:prompts}) and asking participants to think aloud. Again, after seven minutes, the participant's search was stopped by the investigator in case no earlier stopping decision was made.

\textbf{Step VI: Experience with Guided Task.}
The participants filled out the same questionnaire and answered the same interview questions as in the unguided task.

\textbf{Step VII: Demographics.}
Participants were asked if they now would use an LLM for data search in the future.
Then, they filled out a short demographic survey and had the opportunity to comment on the study or to mention anything they could not bring up before. The recording was then stopped.

In total, the study took about an hour to complete for each participant.

\subsection{Used LLMs}\label{sec:llms}

ChatGPT has shown high performance in various scenarios, such as text summarisation, clinical decision support, or question answering~\cite{chatgpt_llama,chatgpt_mistral_llama,sandmann,Seth2023ComparingTE,Shukla,Silva2024ChatGPTVL,wangsa}.
Therefore, we opted to use ChatGPT Scholar GPT, an OpenAI model focused on academic tasks with access to the web (called \textit{ChatGPT} for short) in the Plus version. Additionally, we include Perplexity with a focus on academic with web search enabled (called \textit{Perplexity} for short) as a comparison in our study for its ability to provide references~\cite{Shukla}. Both of these LLMs and settings provide online sources for search results, which we consider important for using LLMs in research scenarios.

\subsection{Study Participants}

We recruited participants for our study mainly by reaching out to 41 personal contacts from various fields who we asked to distribute a call for participation at their institution or in their wider network. In addition, we sent calls for participation to 13 mailing lists in different disciplines and posted the call on social media (Bluesky and Mastodon). We were able to recruit 32 participants from Germany (23 participants), Japan (2), Lithuania (2), USA (2), Canada (1), Singapore (1), and the United Kingdom (1).

Sixteen participants identify as male, 16 as female and their mean age is 36.47 years ($\sigma = 8.22$). Two participants have a bachelor's degree (or equivalent), 12 have a master's degree (or equivalent), 16 have a Ph.D., and 2 participants have a habilitation, a German post-doctoral qualification for becoming a professor. They stem from various research fields. Twenty participants do research in the social sciences, 7 in technology, 3 in arts and humanities, and 1 in life sciences and biomedicine.

When asked about their experience with research data, 9 said they use research data every day, 12 use data more than once a week, 6 use data once a week, and 5 use data one to three times a month. On average, our participants have worked with data for 10.77 years ($\sigma = 6.98$).
We also asked them about their experience with LLM use. Ten said they use LLMs every day, 13 use LLMs more than once a week, 2 one to three times a month, 1 several times a year, 2 said they use LLMs less often than that, and 4 never used LLMs.

While recruiting participants, we screened for their experience in data search and with usage of LLMs. This was to achieve a more or less equal assignment to one of the two LLMs based on their experience. \autoref{tab:experience} indicates participants' experience in data search and with usage of LLMs (low vs. high), based on the more exact information they provided during the interview and survey, and their assigned LLM.

\begin{table}[t]
    \caption{Distribution of participants to LLMs with self-reported experience in data search and with usage of LLMs.}

    \centering
    \tiny
    \begin{tabular}{l|l!{\vrule width 1.5pt}c|c}
        \multicolumn{2}{c!{\vrule width 1.5pt}}{\textbf{Experience}} & \multicolumn{2}{c}{\textbf{LLMs}} \\ 
        \textbf{Data Search} & \textbf{LLM Usage} & \textbf{ChatGPT} & \textbf{Perplexity} \\ \hline
        low & low & 1 & 1 \\
        low & high & 9 & 6 \\ 
        high & low & 3 & 4 \\
        high & high & 3 & 5\\
        
    \end{tabular}
    \label{tab:experience}
\end{table}

\begin{table}[t]
    \caption{Overview of annotated dialogue acts with explanations and examples from the collected data. Upper part describes dialogue acts annotated both for participants and LLMs, lower part describes dialogue acts only relevant for participants' prompts.}
    \centering
    \tiny

    \begin{tabular}{l|p{4.5cm}|p{5.75cm}}
        \textbf{Component} & \textbf{Explanation} & \textbf{Example from data}\\ \hline
        Statement & Descriptive, narrative, personal & \textit{There are several other panel studies from European countries that include questions on gender, gender roles, and family attitudes.} (id=13)\\ 
        Opinion & Often including hedges such as \textit{think, believe, seem, mean} & \textit{I think your sugegstions makes sense but this is not what I am exactly trying to do.} (id=23)\\ 
        Question & Differentiation between \textit{yes/no}, \textit{wh-word} and \textit{multiple yes/no} questions & \textit{How were each of the non-database-wide datasets collected} (id=12)\\ 
        Answer & Follows a yes/no question, differentiation between \textit{yes} and \textit{no} answers & \textit{do you have any supply chain dataset sources?} (id=11)\\ 
        Appreciation & Appreciation or gratefulness& \textit{That's an interesting project you're working on.} (id=11)\\ 
        Apology & Regretful acknowledgement of mistake/failure & \textit{I apologize for the confusion in my previous response.} (id=11) \\ 
        Thanking & Expression of gratitude & \textit{Thanks for pointing that out!} (id=14)\\ 
        Action & Demanding an action done by conversational partner & \textit{Please summarise briefly each dataset and the size of each and originating databases for each dataset} (id=12)\\ 
        Offer & Offering doing an action to the conversational partner & \textit{Let me know if you need access links or further details on a specific dataset.} (id=12)\\ 
        Opening & Conversational opening & \textit{Hi Leslie.} (id=13)\\ 
        Closing & Conversational closing & \textit{Yes, that was my idea, and I think I got some answers} (id=15)\\ 
        Hold & Indicating waiting & \textit{I will now gather relevant datasets. Please hold on.} (id=13)\\ 
        \hline
        Abandon & Stopping the interaction & - stopping the LLM's creation of a response -  (id=19)\\ 
        Repeat phrase & Repeating a previous phrase with little change similar to copy-pasting & \textit{Does the British Election Panel Study contains questions on gender}, \textit{Does the British Household Panel Study contain questions on gender} (id=13)\\ 
    \end{tabular}
    \label{tab:dialogue}
\end{table}

\subsection{Data Processing and Annotation}

We initially created two data files:
\begin{enumerate*}
    \item A hierarchical data file that contains multiple rows per participant and task (unguided and guided) with the submitted queries and received LLM responses (\texttt{LLM-interactions}). In addition we coded further information either manually, for example whether participants defined/assigned a role for themselves or the LLM, or using Python, such as query and response length, the use of emojis, and whether the LLM was addressed by the persona's name. We anonymised the \texttt{LLM-interactions} for publication and will refer to participants by random ids when citing from this file~\cite{ZA8800}. 
    \item A flat data file that contains the responses to the interview questions and the survey questions between the two tasks and at the end of the session (\texttt{Background\_questions}). The survey questions were exported from the survey software LimeSurvey~\cite{LimeSurvey}, cleaned, coded, and labelled using Stata. In particular we coded the five UEQ subscales we used\footnote{We used \texttt{Attractiveness}, \texttt{Efficiency}, \texttt{Perspicuity}, \texttt{Dependability}, and \texttt{Stimulation}.}, which also included reverse coding for some items. The interview questions were also coded and added to the \texttt{Background\_questions}. Furthermore, for some coded information from the \texttt{LLM-interactions} we calculated averages, pivoted them into a flat data structure, and merged them with the \texttt{Background\_questions}~\cite{ZA8800}.
\end{enumerate*}

The \texttt{LLM-interactions} were then annotated to create a third data file (\texttt{Annotations-file})~\cite{ZA8800}. 
One trained annotator labelled all prompts and responses in the \texttt{LLM-interactions} with regard to contained components\footnote{Our study being a written conversation between LLM and participant prevents several dialogue acts, e.g., \texttt{non-verbal}, \texttt{backchannel}, \texttt{self-talk}, and \texttt{3rd party talk}.} from a subset of dialogue acts according to Stolcke et al.~\cite{stolcke-etal-2000-dialogue} (see \autoref{tab:dialogue}).

The data have very few missings. In the \texttt{LLM-file}, responses from one participant-LLM interaction could not be saved due to technical reasons and are therefore not included. In the \texttt{Survey-file} we have very few cases of item-non-response in the UEQ-scales. Respondents who skipped UEQ items are excluded from analyses involving the affected subscales. We deleted one participant's research field to protect their identity as they have a very narrow research topic.

\section{Analysis}

As we strive to tackle our overarching research objective of investigating \textit{the information seeking practices of researchers looking for research data using LLMs}, this first work examines the following aspects focusing on the participants' interactions with the LLMs:

\begin{itemize}
    \item RQ$_1$: What is the observable interaction behaviour between participants and the LLMs?
    \item RQ$_2$: Do the interactions between participants and LLMs show characteristics associated with \textit{humanisation}?
    \item RQ$_3$: Can we observe differences in the interactions between participants from \textit{different user groups}?
\end{itemize}

\subsection{RQ$_1$: Participants' information behaviour with LLMs}

To answer our first research question, we examine quantitative properties as well as a high-level qualitative classification of participants' queries and LLMs' responses.
We consider these aspects: 
\begin{enumerate*}
    \item How long are interaction \textit{sessions}, how does the \textit{length of prompts} and responses change throughout sessions?
    \item How did the participants formulate their prompts on a continuum from keyword queries to phrases or a concise natural language query?
    \item Did participants search for \textit{data or papers}?
\end{enumerate*}




\subsubsection*{Sessions and length of prompts.}
A quantitative analysis of queries and responses gives us first insights into whether there are considerable differences between usage of the different LLMs and tasks.
We collected 264 pairs of participant-issued queries with LLMs' responses\footnote{We do not consider our persona-prompting query-response pairs in the guided task as part of the conversations.}. 127 pairs stem from the unguided and 137 pairs stem from the guided task. The number of interactions is considerably higher for Perplexity (68 in total with 4.25 pairs on average for Perplexity, 59 in total with 3.69 pairs on average for ChatGPT), which is even more pronounced in the guided scenario (79 in total with 4.94 conversation pairs per session on average for Perplexity, 58 in total with 3.62 pairs on average for ChatGPT).

\begin{figure}[t]
\includegraphics[scale=0.5]{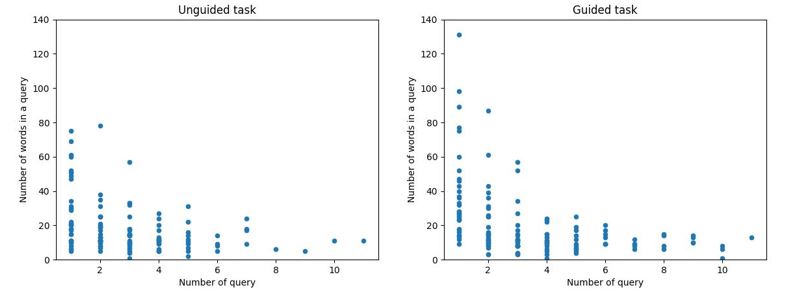}
\centering
\caption{Length of queries over the course of all sessions in both tasks.}
\label{fig:sessionlength}
\end{figure}

\autoref{tab:sessionlength} gives an overview of the session characteristics and differences between the unguided and guided task. We did not find significant differences in the number of queries that participants submitted between tasks. The average length (in number of words) of queries and of responses differ significantly between the two tasks and we see in \autoref{fig:sessionlength} that the query lengths decrease over the course of the sessions. In the unguided task, the queries were shorter than in the guided task. However, the LLM provided longer responses in the unguided task vs. the guided task. The noun to total words ratio in participants' queries gives an indication of whether a keyword search is used rather than phrases (see also next subsection). We find that, in relation to other words, more nouns are used in the unguided task than in the guided one. The number of emojis used in responses from ChatGPT (see the subsection \textbf{Emojis} in \autoref{sec:rq2_humanisation}) does not differ significantly between tasks. 

\begin{table}[t]
    \caption{Session characteristics for the unguided and guided tasks and paired t-test between the tasks.}

    \centering
    \tiny
    \begin{tabular}{l|rr|rr|lr}
	\ & \multicolumn{2}{c|}{\textbf{Unguided}} & \multicolumn{2}{c|}{\textbf{Guided}} & \multicolumn{2}{c}{\textbf{Paired t-test}} \\ \hline
    Average number of queries & 3.9688  & (0.3674) & 4.2813 & (0.4693) &-1.0000  & (0.3251) \\
	Average query length (words) & 21.7663 & (2.4539)  & 33.5357 & (2.4456) & -5.5272*** & (0.0000)\\
    Average response length (words) & 303.5189 & (21.9460) & 207.3831 & (10.2393) & \hspace{2.5pt}4.2321*** & (0.0002) \\
    Average noun to total words ratio & 0.3608 & (0.0151) & 0.2925 & (0.0094) & \hspace{2.5pt}3.8635***  & (0.0005)\\
    Average number of emojis & 3.2272 & (1.4251) & 2.8620  & (0.9237) & \hspace{2.5pt}0.2253 & (0.8248) \\

    \end{tabular}

    Notes. Mean (with standard error) for session characteristics and two data search tasks (unguided and guided), t-statistic (and two-tailed p-value) for paired t-test performed between two search tasks, *** $p < 0.001$, $N = 32$.
    \label{tab:sessionlength}
\end{table}

\subsubsection*{Prompt types.}
The formulation of prompts could differ between users. Inexperienced users might try to use the LLM as they would use a search engine while more experienced users might be utilising LLMs' capabilities to operate with natural language queries.
Here, we again analyse the manually annotated prompts issued by participants in the unguided and guided tasks. 
Over both tasks, we encounter similar prompt formulation patterns. Participants mostly issued their prompts as concise natural language queries (119 for the guided task, 132 for the unguided task). For the (un)guided task, 2 (1) queries were keyword queries, 1 (0) was a mixture of keyword query and phrase, 3 (2) were formulated as phrases but not as natural language, and 2 (2) prompts were formulated as queries which were more concise than phrases but not quite natural language.

\subsubsection*{Data or papers.}
Strategies for paper search do not transfer directly to data search~\cite{10.1145/2232817.2232822}. So, distinguishing between searching for data and papers in our search tasks gives us insight into the primary objective of participants and the opportunity to compare employed strategies.
Here, we consider the manually annotated prompts issued by participants belonging to the different sessions during the unguided and guided tasks.
In 29 unguided and 29 guided sessions data was explicitly searched for, in 3 unguided and 8 guided sessions papers were searched for. Two unguided and 6 guided sessions included inquiries for both data and papers.
27 participants did search for data in both tasks, while only one participant searched for papers in both scenarios.

\subsubsection*{Discussion.}
Queries are shorter in the unguided task and the noun to total words ratio is larger. This indicates that shorter searches or potentially keyword searches were more likely to occur during the unguided tasks. In the unguided tasks, the LLMs provided longer responses. Still, the manual investigation of queries showed almost all queries being natural language queries.
Over both tasks we encountered fewer query-response pairs in case ChatGPT was used. The number of pairs was stable over both tasks for ChatGPT but increased considerably for the guided task when using Perplexity.

Nearly all participants asked directly for data. Only very few did try to find papers using data they are looking for.

Regarding the interaction behaviour between participants and LLMs we conclude: \textbf{Prompts are issued mostly as natural language queries, especially during the guided task, and participants did mainly search for data directly.}

\subsection{RQ$_2$: Humanisation}
\label{sec:rq2_humanisation}

To answer our second research question targeting aspects of humanisation in the interactions, we strive to find out if participants use the LLM as a tool or treat it as a conversational partner and if the behaviour differs in the setting where the LLM has been primed to act as a persona. If the (primed) LLM would be treated as an equal conversational partner, this could hint at LLMs' suitability to serve as a stand-in for "data talk"~\cite{DBLP:conf/asist/Yoon17} in case a human would not be available. 
We examine the following four aspects:
\begin{enumerate*}
    \item Do participants' and LLMs' utilised \textit{dialogue acts} differ in their interactions?
    \item Do participants define their \textit{own role} and/or a \textit{role for the LLM} when searching for data?
    \item Do LLMs and participants use emojis?
    \item Does participants' \textit{user experience} differ between using the LLM intuitively and using the LLM with the prompt asking it to behave as a persona?
\end{enumerate*}

\subsubsection*{Dialogue acts.}
Dialogue acts, especially social ones such as apologies or appreciation, might signal anthropomorphic traits. If participants and LLMs show similar usage of dialogue acts, this could hint at humanisation of LLMs.
Here, we consider the manually annotated prompts issued by participants and the LLMs' responses in the unguided and guided tasks.
\autoref{tab:dialogue_acts} gives an overview of dialogue acts found in the sessions and single queries/responses per task for both used LLMs in combination.\footnote{Detailed results can be found in the \href{https://github.com/kreutzch/TPDL25_DataDiscoveryUsingLLMs/blob/main/Online_Appendix-Data_Discovery_using_LLMs.pdf}{online appendix on GitHub}.}

\begin{table}[t]
    \caption{Number of sessions (out of 32) in which dialogue acts were present for prompts issued by participants and responses from LLMs for the unguided and guided tasks. Numbers in brackets indicate the percentage of queries/responses in which a dialogue act was found rounded to two decimal places.}

    \tiny
    \centering
    \begin{tabular}{l|rr|rr|rr|rr}
        \textbf{Actor} & \multicolumn{4}{c|}{\textbf{Participant}} & \multicolumn{4}{c}{\textbf{LLM}} \\
        \textbf{Task} & \multicolumn{2}{c|}{\textbf{Unguided}} & \multicolumn{2}{c|}{\textbf{Guided}} &  \multicolumn{2}{c|}{\textbf{Unguided}} & \multicolumn{2}{c}{\textbf{Guided}}\\ \hline

Statement & 
20 &(32.28\%) & 29 &(37.96\%)
 & 
32 &(99.21\%) & 32 &(97.81\%)
\\
Opinion & 
1 &(0.79\%) & 5 &(3.65\%)
 & 
1 &(0.79\%) & 3 &(3.65\%)
\\
Question & 
28 &(55.91\%) & 32 &(76.64\%)
 & 
12 &(29.13\%) & 17 &(34.31\%)
\\
Answers & 
3 &(3.94\%) & 5 &(5.84\%)
 & 
16 &(22.83\%) & 23 &(28.47\%)
\\
Appreciation & 
0 &(0.0\%) & 2 &(1.46\%)
 & 
1 &(0.79\%) & 13 &(12.41\%)
\\
Apology & 
0 &(0.0\%) & 1 &(0.73\%)
 & 
2 &(3.15\%) & 5 &(3.65\%)
\\
Thanking & 
2 &(1.57\%) & 10 &(8.76\%)
 & 
1 &(0.79\%) & 1 &(0.73\%)
\\
Action & 
26 &(49.61\%) & 22 &(27.01\%)
 & 
0 &(0.0\%) & 0 &(0.0\%)
\\
Offer & 
0 &(0.0\%) & 0 &(0.0\%)
 & 
17 &(40.94\%) & 25 &(55.47\%)
\\
Opening & 
2 &(2.36\%) & 21 &(16.06\%)
 & 
0 &(0.0\%) & 22 &(18.98\%)
\\
Closing & 
0 &(0.0\%) & 2 &(1.46\%)
 & 
0 &(0.0\%) & 1 &(0.73\%)
\\
Hold & 
0 &(0.0\%) & 0 &(0.0\%)
 & 
1 &(0.79\%) & 4 &(2.92\%)
\\ \hline
Repeat phrase & 
5 &(7.09\%) & 4 &(3.65\%)
 \\
Abandon & 
3 &(2.36\%) & 1 &(0.73\%)
 \\
    \end{tabular}
    \label{tab:dialogue_acts}
\end{table}

In general, responses from LLMs contain \texttt{statements}. Prompts issued by participants contain \texttt{statements} in about one third of the cases. 

Almost all sessions contain \texttt{questions} posed by participants. About half of the queries in the unguided task contain a \texttt{question} independent of which LLM is interacted with. This number increases to 76.64\% in the guided task.
Around half of the sessions contain \texttt{questions} posed by an LLM but only one third of responses contain  \texttt{questions} which means that LLMs did pose questions sparsely. Looking into posed questions by LLMs shows that they mostly use \texttt{questions} to \texttt{offer} possibilities for next steps, e.g., \textit{Would you like help \textbf{accessing specific data} from these surveys for Lithuania and the Baltic region?} (id = 15).

Calls to \texttt{action} are only performed by participants, never by an LLM independent of setting.
Similarly, \texttt{offers} are only ever issued by LLMs, never by participants independent of setting. Perplexity does issue considerably fewer \texttt{offers} in responses (2.94\% in the unguided and 22.78\% in the guided task) compared to ChatGPT where we encountered an \texttt{offer} in almost all responses (84.75\% in the unguided and 100\% in the guided task).

In prompts, the \texttt{questions} issued by participants can oftentimes be considered as calls to \texttt{action} also, e.g., \textit{Could you update the list of sources accordingly?} (id = 30).
If such a call to \texttt{action} is formulated as a \texttt{question}, the LLM often does not \texttt{answer} the question but presents the results fitting the required action immediately, e.g., \textit{Based on your investigation into mapping, comparing, and evaluating disciplinary classification systems, I can suggest several relevant articles and search queries to help you find comprehensive literature for your research.} (id = 40).
In responses given by the LLMs, \texttt{offers} are usually not spelled out explicitly (e.g., \textit{Let me know if you need access links or further details on a specific dataset. \includegraphics[height=1em]{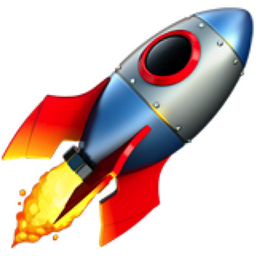} } (id = 12), or can be contained in \texttt{questions} as stated before.

In the guided task, we encountered acts of \texttt{appreciation} coming from an LLM in around one third of the sessions, while we only encountered a single session with such an \texttt{appreciation} in the unguided setting.

In the guided setting, we encounter around 20\% of responses from LLMs containing \texttt{openings}, in the unguided setting we do not encounter a single one coming from LLMs. Similarly, in the unguided setting, we encounter none or very few \texttt{openings}, \texttt{thankings}, \texttt{apologies}, and \texttt{appreciations} stemming from participants, while these numbers slightly increase for the guided setting.
Participants do issue a few \texttt{thankings}, e.g., as an empty phrase instead of out of gratitude \textit{Hi! Thanks for helping me with this task.} (id = 30) or at the end of a call to action \textit{Do you know about some secondary literature which would help me to start? Because I do not have much of a background in this topic. Thanks} (id = 28).

When analysing \texttt{questions} and responses more closely, we found that 68 queries by participants posed wh-word questions (e.g., \textit{how}, \textit{what}, or \textit{which}) to LLMs and 108 yes/no questions. Of these yes/no questions, 65 were explicitly answered by the LLM immediately.
LLMs issued wh-word \texttt{questions} in 7 cases and yes/no questions in 77 responses. Participants \texttt{answered} these yes/no questions 11 times.
As stated before, \texttt{questions} posed by the LLMs were mainly \texttt{offering} further assistance which participants seemed to mostly ignore.

\subsubsection*{Definition of role.} 
Defining one's role in a conversation can serve as social positioning or establishing context. For effectively prompting LLMs, defining a persona or role is an established technique~\cite{white2023promptpatterncatalogenhance}. As this prompting technique hints that participants do not consider the LLM a conversational partner but rather a tool, we here consider queries issued by participants.
Eight participants started their unguided data search by clarifying their own role, for example by stating, \textit{As a researcher I am interested in ...} (id = 20). In the same unguided task, only 4 participants assigned a specific role to the LLM, for example, \textit{You are an expert computer science researcher based in ...} (id = 37). Of those 4 participants, 3 had defined their own role and a role for the LLM, 1 participant defined a role only for the LLM but not for themself. In the guided task (here the LLM already had an assigned role), 8 participants also defined their own role. Three participants stated their own role in both tasks, 5 researchers each stated their own role only in one of the tasks.

\begin{table}[t]
    \caption{Top 10 emojis (and frequency) in ChatGPT's responses in both tasks.}

    \centering
    \tiny
\begin{tabular}{l|rr|rr|rr|rr|rr|rr|rr|rr|rr|rr}
\textbf{Rank}     & \multicolumn{2}{c|}{\textbf{1}} & \multicolumn{2}{c|}{\textbf{2}} & \multicolumn{2}{c|}{\textbf{3}} & \multicolumn{2}{c|}{\textbf{4}} & \multicolumn{2}{c|}{\textbf{5}} & \multicolumn{2}{c|}{\textbf{6}} & \multicolumn{2}{c|}{\textbf{7}} & \multicolumn{2}{c|}{\textbf{8}} & \multicolumn{2}{c|}{\textbf{9}} & \multicolumn{2}{c}{\textbf{10}} \\ \hline
\textbf{Unguided} &     \includegraphics[height=1em]{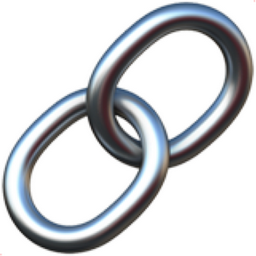}         & (37)            &         \includegraphics[height=1em]{figures/rocket.png}      & (24)            &        \includegraphics[height=1em]{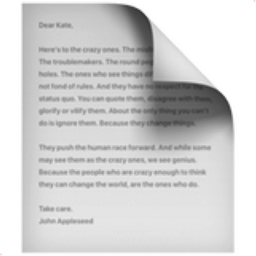}      & (12)            &      \includegraphics[height=1em]{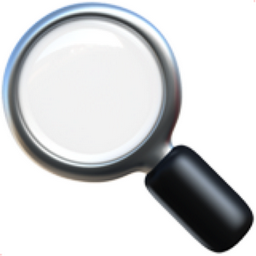}        & (11)            &       \includegraphics[height=1em]{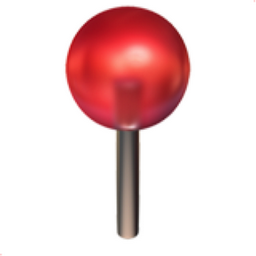}       & (11)            &        \includegraphics[height=1em]{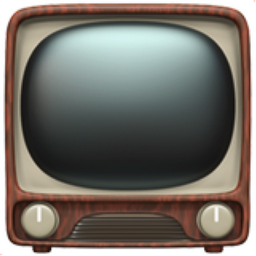}      & (11)            &       \includegraphics[height=1em]{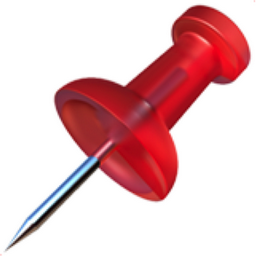}       & (10)            &       \includegraphics[height=1em]{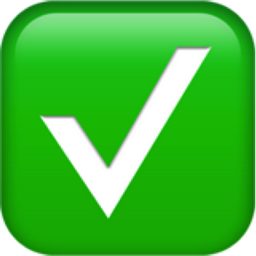}       & (9)             &      \includegraphics[height=1em]{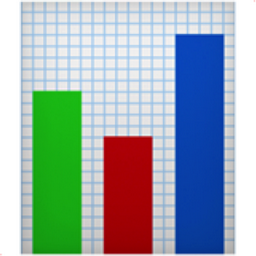}        & (5)             &      \includegraphics[height=1em]{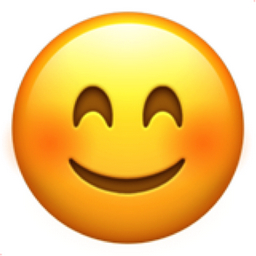}         & (4)             \\
\textbf{Guided}   &       \includegraphics[height=1em]{figures/page-facing-up.png}       & (63)            &      \includegraphics[height=1em]{figures/pushpin.png}        & (34)            &      \includegraphics[height=1em]{figures/rocket.png}         & (27)            &       \includegraphics[height=1em]{figures/link}       & (27)            &        \includegraphics[height=1em]{figures/bar-chart.png}      & (11)            &     \includegraphics[height=1em]{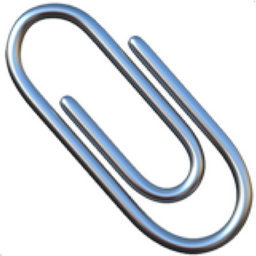}         & (11)            &      \includegraphics[height=1em]{figures/white-heavy-check-mark.png}        & (10)            &        \includegraphics[height=1em]{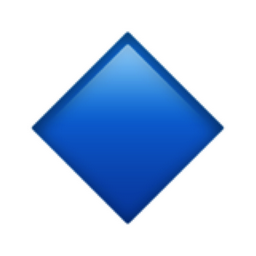}      & (7)             &        \includegraphics[height=1em]{figures/smiling-face-with-smiling-eyes.png}      & (5)             &       \includegraphics[height=1em]{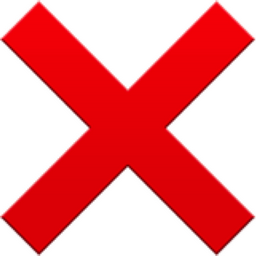}        & (4)            
\end{tabular}
    \label{tab:emoji}

\end{table}

\subsubsection*{Emojis.}
Written conversations between humans contain emojis to provide additional social cues~\cite{emojisforconversations}. Using emojis in chatbots can increase the perceived warmth (not competence) and increase user satisfaction, especially in hedonic scenarios~\cite{YU2024102071}.
So for this aspect we investigate participants' queries and LLMs' outputs in the unguided and guided tasks. Participants and Perplexity never used any emoji. ChatGPT used 264 emojis in total and 35 unique ones in its responses. In the unguided task, ChatGPT used fewer emojis (127) than in the guided one (137), but the choice of emojis was more varied in the unguided scenario: 32 unique emojis in the unguided vs. 14 in the guided task. \autoref{tab:emoji} displays the ten most common emojis in the ChatGPT responses.
Most emojis used by ChatGPT were used to structure its findings. For example, a found data source was presented with \includegraphics[height=1em]{figures/page-facing-up.png} preceding a data or a study title, \includegraphics[height=1em]{figures/bar-chart.png} referred to the actual dataset, \includegraphics[height=1em]{figures/pushpin.png} was followed by a short summary of the paper or study, and \includegraphics[height=1em]{figures/link} or a \includegraphics[height=1em]{figures/left-pointing-magnifying-glass.png} referred to further sources or information about the finding. includegraphics[height=1em]{figures/television.png} (used 11 times in the unguided task) was solely used for one participant who was doing research on telenovelas. \includegraphics[height=1em]{figures/rocket.png} was very often used at the end of the LLM response conveying an informal sense of excitement, success, and ambition.


\subsubsection*{User experience.}
We examine whether there are differences in the participants' user experience when they use the LLM freely compared to when the LLM is primed to behave as a persona.
This aspect considers the participants' evaluation of their user experience in the unguided and guided tasks. Using the User Experience Questionnaire (UEQ)~\cite{UEQ}, participants rated their search mainly positive and the ratings do not differ significantly between both tasks (see \autoref{tab:userexp}). The lowest mean scores were given for the \texttt{Dependability} subscale. This subscale measures whether users feel in control and whether the interaction with a tool is secure and predictable. The highest mean scores were reached for the \texttt{Perspicuity} subscale which measures whether it is easy to get familiar with a tool and to learn how to use it.

For the NASA Task Load Index \cite{hart1988development} the mean scores are mostly close to the middle scale point of 3.5 (see \autoref{tab:userexp}). Only \textit{Physical demand} and \textit{Frustration} have lower scores than 3.0 in both tasks. \texttt{Performance} is rated highest. We did not find a significant difference between the unguided and the guided task.

\begin{table}[t]
    \caption{Participants' user experience in the unguided and guided tasks.}

    \centering
    \tiny
    \begin{tabular}{l|rr|rr|rr|r}
	\ & \multicolumn{2}{c|}{\textbf{Unguided}} & \multicolumn{2}{c|}{\textbf{Guided}} & \multicolumn{2}{c|}{\textbf{Paired t-test}} & \textbf{N} \\ 
    \hline
    
    \multicolumn{8}{l}{\textbf{User experience questionnaire (UEQ)}} \\
    Attractiveness & 1.0108 & (0.1884) & 1.1398 & (0.2038) & -0.6329 & (0.5316) & 31 \\
    Efficiency & 1.2500 & (0.1912) & 1.1935 & (0.2208) & -0.2670 & (0.7965) & 31 \\
    Perspicuity & 1.7734 & (0.1376) & 1.5938 & (0.1633) & 1.0738 & (0.2912) & 32 \\
    Dependability & 0.8214 & (0.1329) & 0.9018 & (0.1856) & -0.5068 & (0.6164) & 28 \\
    Stimulation & 1.0391 & (0.1935) & 1.0938 & (0.2213) & -0.2297 & (0.8198)& 32 \\

    \multicolumn{8}{l}{\textbf{NASA Task Load Index}} \\
    Mental demand & 3.5938 & (0.2651) & 3.5000 & (0.2731) & 0.3728 & (0.7118) &  32 \\
    Physical demand & 1.6875 & (0.1928) & 1.9063 & (0.2349) & -1.0450 & (0.3041) & 32 \\
    Temporal demand & 3.6875 & (0.2673) & 3.1250 & (0.2901) & 1.8688 & (0.0711) & 32 \\
    Performance & 3.8125 & (0.3189) & 3.5000 & (0.2910) & 0.8808 & (0.3852) & 32 \\
    Effort & 3.3750 & (0.2408) & 2.9375 & (0.2150) & 1.4222 & (0.1650) & 32 \\
    Frustration & 2.9375 & (0.3076) & 2.9375 & (0.3010) & (0.0000) & 1.0000 & 32 \\
    \end{tabular}

    Notes. Mean with (standard error) for five UEQ scales \cite{UEQ} and the NASA Task Load Index \cite{hart1988development} and two data search scenarios (unguided and guided), t-statistic and two-tailed p-value for paired t-test performed between two data search scenarios, number of observations. A Wilcoxon Signed Rank was performed for Efficiency (UEQ) and the z-statistic and exact p-value reported. UEQ scales range from -3 to + 3, NASA Task Load Index scales range from 1 to 7.
    \label{tab:userexp}
\end{table}

\subsubsection*{Discussion.}

For data search in a typical digital library setting, it has been observed that users re-issued queries in almost 70\% of cases~\cite{DBLP:conf/ercimdl/CarevicR020}. While our study found few repeated phrases in the conversational search setting, our sessions were typically focused on finding datasets for one pre-defined information need.
Considering the dialogue acts, we see differences between participants and LLMs for both tasks in actions, offers, questions, answers and more social ones such as appreciation. From this we derive that participants seem to not treat the LLM as a conversational partner but rather as a tool.

Few participants defined their and the LLM's roles. We did not encounter differences between the two tasks so the pre-definition of a persona does not seem to change our participants' behaviour in this regard.

The absence of emojis in participants' queries could stem from the fact, that they were not familiar with using emojis on a computer keyboard. 
Contrasting Perplexity, ChatGPT incorporates emojis into its responses, mainly for structuring purposes, but some (\includegraphics[height=1em]{figures/rocket.png}) also convey non-obvious informal meaning.
Emojis can be used to match an interlocutor's interaction style~\cite{CAVALHEIRO2023102023} but our participants did not start using emojis after ChatGPT used them and ChatGPT did not stop using emojis after participants did not respond to them.

We did not find significant differences in the user experience between the unguided and guided task. This is surprising, as the LLMs showed different conversational actions but this seems to not have affected users' perception.

Regarding aspects of humanisation in the interactions between participants and LLMs we conclude: \textbf{While LLMs seem to hold a conversation and strive to appear human-like to an extent, participants do not appear to recognise them as equal conversational partners. Instead they rather use them as tools. Prompting an LLM to behave as a persona does change its way of interacting but does not influence users' experience using it.}

\subsection{RQ$_3$: Differences between groups of participants}

We expect that searching for data with an LLM is affected by participants' previous experience with research data and whether LLMs are already being used while doing research. To address our third research question targeting differences between experience groups, we consider the following two aspects:
\begin{enumerate*}
    \item Do interactions with the LLM differ between participants experienced and inexperienced with \textit{searching for or working with data}?
    \item Do interactions with the LLM differ between participants experienced and inexperienced with \textit{using LLM}?
    \item Does the user experience differ between participants experienced and inexperienced with \textit{searching for or working with data}?
    \item Does the user experience differ between participants experienced and inexperienced with \textit{using LLMs}?
\end{enumerate*}

During the interview before the data search tasks and in the survey at the end of the session, we asked participants to assess their experience in using LLMs and with searching and working with data using LLMs. We divided participants into two groups (high/low) each based on their responses. Our detailed results can be found in our \href{https://github.com/kreutzch/TPDL25_DataDiscoveryUsingLLMs/blob/main/Online_Appendix-Data_Discovery_using_LLMs.pdf}{online appendix on GitHub}.

\subsubsection*{Interactions with LLMs and experience groups.}
We see few differences in session characteristics between experience groups: Participants with high experience in working with data used more queries than participants with low experience in the unguided task. Also in the unguided task, participants with high LLM experience formulated longer queries than participants with low LLM experience, while in the guided task, they used fewer queries than participants with low LLM experience.

\subsubsection*{User experience and experience groups.}
Participants with high data search experience perceived the data search with an LLM less efficient in both tasks and they perceived higher temporal demand in the unguided task compared to participants with high experience. In the guided task they also found the LLM search less stimulating. Participants with high LLM experience rated all five UEQ subscales in the guided task higher than participants with low LLM experience. However, also for the guided task and using the NASA Task Load Index, they rated the LLMs' performance lower than their peers with low experience.

\subsubsection*{Discussion.}
In interacting with the LLM, we especially saw differences for participants who used LLMs frequently and are therefore experienced in using LLMs. Compared to participants with less experience, they formulated significantly longer queries in the unguided task and significantly fewer queries in the guided task. At the same time they perceived higher \texttt{Attractiveness}, \texttt{Efficiency}, \texttt{Perspicuity}, \texttt{Dependability}, and \texttt{Stimulation} when working on the guided task. Participants with low LLM experience, on the other hand, rated the LLMs' performance higher. They might have been positively surprised by how well the LLMs work without prior experience with the validity of LLMs' results. That participants with low experience in data search reported lower temporal demand in the unguided task, compared to participants with higher experience, could be a sign that this group does not perceive the task as time-critical because they did not grasp the exact data requirements and were not aware of potential steps in the data search that might need to follow.

Regarding differences between experience groups we conclude: \textbf{Participants with prior LLM experience formulate longer queries and, when the LLM is prompted to act as a persona, rate their user experience more positively. Participants experienced in data search rate the data search with an LLM as less efficient than participants with low data search experience.}

\subsection{Discussion and Limitations}



LLMs strive to appear friendly, helpful and approachable, especially in the guided scenario. 
These efforts do not seem to be reciprocated by participants, even though they issue natural language queries, partially respond to LLMs, and keep the conversation flowing without repetition, so they engage in the conversation.
Still, LLMs are used as tools rather than seen as equal conversational partners. 
Therefore, we assume they do not naturally replace the typical data talk with a human colleague.
Prompting the LLM to behave as a persona (as seen in the guided task) does change the LLMs' as well as participants' behaviour - the queries are longer and queries appear to become even more natural. Participants who are experienced in data search find the LLM data search generally less efficient. But when prior experience with LLMs exists, users' report higher satisfaction with the tool when prompted to behave as a persona.
While ChatGPT was shown to have problems in explaining scientific publications and exploring data from scientific databases~\cite{AlSagriForthcoming-ALSCOG}, we did not specifically encounter these problems with our chosen model and setting.

Our study comes with some \textbf{limitations}: 
\begin{enumerate*}
    \item We only investigated two settings and two LLMs in an explorative study. Broader settings and also more targeted hypotheses will be needed to expand our knowledge in this area.
    \item While 32 participants from different domains and with different expertise levels is remarkable for this type of study, further and larger user studies representing different domains equally are needed. 
\end{enumerate*}
These limitations lead to a limited generalisability of the conclusions resulting from our analyses.
We consciously refrain from considering or judging the correctness of results provided by the LLMs. Fabrication remains an open problem with LLMs~\cite{Huang2023ASO,minaee2025largelanguagemodelssurvey}.

\section{Conclusion}

We presented our user study focusing on researchers' information interaction behaviour using LLMs for data search with 32 participants from different domains. We quantitatively and qualitatively analysed our collected interaction data and found the following:
\begin{enumerate*}
    \item LLMs seemed to strive to connect with our participants or please them over both tasks. Still, users did not accept them as equal conversational partners which leads to the conclusion of LLMs not being a suitable replacement for a human interlocutor in data talk purposes.
    \item Prompting the LLMs to behave as the colleague of a participant changed the LLMs' and participants' behaviours. Participants experienced with LLMs reported better user experience with the pre-prompted sessions compared to the unguided setting. 
\end{enumerate*}

Future work will continue to investigate the plethora of data we collected in our user study. 
First, we want to focus on users' expression of information needs during the conversations with LLMs.
We will do a more in-depth analysis of the dialogue acts in combination with the think-aloud data to extract more abstract data search patterns. These patterns will then also be compared to participants' usual data search techniques. 
Another line of work will focus on examining the encountered differences between ChatGPT and Perplexity and the effects on users' behaviour.
Additionally, we intend to investigate participants' information experience more. We will for example investigate, if personas are accepted by participants or only tolerated.

%
%
\bibliographystyle{splncs04}
\bibliography{bibliography}

\end{document}